\documentclass[]{spie}  

\usepackage{amsmath,amsfonts,amssymb}
\usepackage{graphicx}
\usepackage[colorlinks=true, allcolors=blue]{hyperref}
\usepackage{siunitx} 
\usepackage{ulem}

\title{Grazing incidence X-ray fluorescence based characterization of nanostructures for element sensitive profile reconstruction}

\author[a]{Anna Andrle}
\author[a]{Philipp H\"{o}nicke}
\author[b]{Philipp-Immanuel Schneider}
\author[a]{Yves Kayser}
\author[b]{Martin Hammerschmidt}
\author[b,c]{Sven Burger}
\author[a]{Frank Scholze}
\author[a]{Burkhard Beckhoff}
\author[a]{Victor Soltwisch}
\affil[a]{Physikalisch-Technische Bundesanstalt, Abbestrasse 2-12, 10587 Berlin, Germany}
\affil[b]{JCMwave GmbH, Bolivarallee 22, 14050 Berlin, Germany}
\affil[c]{Zuse Institute Berlin, Takustrasse 7, 14195 Berlin, Germany}

\authorinfo{Further author information: (Send correspondence to Anna Andrle)\\A. Andrle: E-mail: anna.andrle@ptb.de, Telephone: +49 30 3481 7148}

\begin{document} 
\maketitle

\begin{abstract}
For the reliable fabrication of the current and next generation of nanostructures it is essential to be able to determine their material composition and dimensional parameters. Using the grazing incidence X-ray fluoresence technique, which is taking advantage of the X-ray standing wave field effect, nanostructures can be investigated with a high sensitivity with respect to the structural and elemental composition. This is demonstrated using lamellar gratings made of Si$_3$N$_4$. Rigorous field simulations obtained from a Maxwell solver based on the finite element method allow to determine the spatial distribution of elemental species and the geometrical shape with sub-nm resolution. 
The increasing complexity of nanostructures and demanded sensitivity for small changes quickly turn the curse of dimensionality for numerical simulation into a problem which can no longer be solved rationally even with massive parallelisation. New optimization schemes, e.g. machine learning, are required to satisfy the metrological requirements. We present reconstruction results obtained with a Bayesian optimization approach to reduce the computational effort.
\end{abstract}


\section{INTRODUCTION}
\label{sec:intro}  


New generations of nanostructures are mostly characterized by a decrease in dimension and an increase in complexity in the composition. For example, in the semiconductor industry this leads to the integration of complex 2D and 3D structures with features in the single-digit nanometer regime \cite{S.Natarajan2014,markov_limits_2014}. The performance of these structures depends on the dimensional parameters and the 3D element compositions. New measurement methods are required that can measure nanostructures quickly and non-destructively. 
Small angle X-ray scattering (GISAXS) \cite{levine_grazing-incidence_1989} or grazing-incidence X-ray fluorescence (GIXRF) \cite{D.K.G.DeBoer1995,P.Hoenicke2015,JAAS_2012, Soltwisch2018a} can probe large areas of the sample with an adequate sensitivity to the dimensional and analytical parameters of the structures \cite{PhysRevLett.94.145504,hofmann_grazing_2009,rueda_grazing-incidence_2012,wernecke_direct_2012-1,gollmer_fabrication_2014,V.Soltwisch2016}. The emitted fluorescence radiation is element specific and thus its angular distribution in the GIXRF is depending on the elemental composition of the structure. 

In GIXRF, the incident angle $\theta_i$ between the X-ray beam and sample surface is typically varied around the critical angle $\theta_c$ for total external reflection. On flat samples, the interference between the incoming beam and the reflected beam results in an X-ray standing wave (XSW) field \cite{Bedzyk_1989,Golovchenko1982}, which can strongly modulate the intensity distribution above and below the reflecting surface depending on the specific layer structure. The intensity modulation inside the XSW field is correlated with the incident angle and the wavelength, and significantly impacts the X-ray fluorescence intensity of an atom depending on its position within the XSW. Performing GIXRF angular scans thus provides information about the in-depth distribution of any probed element within the sample \cite{P.Hoenicke2015,JAAS_2012,P.Hoenicke2009}. 

To model GIXRF angular profiles, an accurate calculation of the XSW intensity is essential. For a 1D system (for example a stratified layer stack) the recursive matrix formalism developed by Parratt \cite{Parratt1954} is often used and implemented in various software packages such as IMD \cite{Windt1998} and XSWini \cite{Pollakowski_2015}. This formalism is rather fast and is an ideal candidate for layered systems. However, if well-ordered 2D or even complex 3D structures are present, these software packages are no longer capable of calculating the XSW\cite{M.Dialameh2017, Soltwisch2018a}. Thus for a GIXRF-based characterization of regularly ordered 2D or 3D nanostructures, which are more relevant to fields such as the semiconductor industry, a novel calculation scheme is required for the XSW field, or in general for the near-field distribution.

Maxwell solvers based on the finite-element method (FEM) are suited for the computation of the electric near-field distribution (or in GIXRF terminology, the local excitation condition) within periodic arrangements of surface structures. They can thus contribute to the simulation and interpretation of GIXRF measurement data of such structures in order to derive the dimensional parameters of the structures as well as information about their elemental composition. Similar studies in the optical spectral range have demonstrated the potential of the {finite-element} method \cite{barth_2017}. Expanding this approach to include the X-ray spectral range is challenging due to the fact that the finite-element discretization of the computational domain necessary for this approach depends on the wavelength of the incoming plane wave in order to ensure the numerical precision of the calculated electric-field distribution. For incident radiation with wavelengths in the nm or sub-nm range and domain sizes of several 100 nm, this seems to be only possible with a high computational effort, at first glance. However, the orientation of the wave vector with respect to the geometrical layout of the sample defines the accessible numerical precision within a reasonable computation time depending on the actual goal of the simulations.
In this work, we use the flexibility and potential of the finite-element method (FEM) for the characterization of a lithographically structured silicon nitride Si$_3$N$_4$ lamellar grating on a silicon substrate.

Reconstructions of nanostructures based on modern statistical methods, including uncertainity estimations, still require function evaluations in the range of $>10^6$. The computation time per function evaluation takes easily several minutes if complex nanostructures or short wavelengths are considered. This quickly results in a problem which can no longer be solved rationally even with massive parallelisation. Thus, new optimization schemes, e.g. machine learning, are required to satisfy the metrological requirements. Here, we study the performance of Bayesian optimization (BO), a method that learns a probabilistic model of the objective function during optimization. We compare the reconstruction results with those obtained by particle swarm optimization (PSO), a conventional heuristic optimization method.


\section{Experimental Details}

The nanograting used here consists of silicon nitride on a silicon substrate. A positive resist, ZEP520A (organic polymer) was spin-coated onto the silicon nitrite. The resist was exposed with electron beam lithography (Vistec EBPG5000+ electron beam recorder, electron acceleration voltage of 100 kV) at the Helmholtz-Zentrum Berlin. After resist development, the lattice was etched with CHF$_3$ into the Si$_3$N$_4$ layer by reactive ion etching. Finally, the remaining resist was removed by an oxygen plasma treatment. The nominal pitch of the grid is $p = 100 \si{\nm}$. A cross-section of a whitness sample is shown in Fig.~\ref{fig:fem_mesh}. The line height depends on the initial thickness of the Si$_3$N$_4$ layer and should therefore be about $h = 90 \si{\nm}$. The line width was set to $w = 40 \si{\nm}$. The total area of the grating is 1 mm x 15 mm. The lines are parallel to the long edge.

The measurements were carried out in the PTB laboratory at the BESSY II electron storage ring at the plane-grating monochromator (PGM) beamline \cite{F.Senf1998} for undulator radiation \cite{B.Beckhoff2009c}. The sample was mounted in a ultrahigh-vacuum (UHV) measurement chamber \cite{J.Lubeck2013}. A 9-axis manipulator provides the possibility to align the sample. The incident angle $\theta_i$ between the X-ray beam and the sample surface can be aligned with an uncertainty below 0.01$^{\circ}$. The azimuthal incidence angle $\varphi_j = 0^{\circ}$ defines the position where the lines are parallel to the plane of the incoming beam (conical diffraction). The mirror symmetry of the grating lines allows the alignment of $\varphi_j$ directly with the GIXRF signal, because the signal depends on $\varphi_j$.
The fluorescence signals are recorded with a silicon drift detector (SDD), which is calibrated with respect to its detector response functions and detection efficiency \cite{F.Scholze2009}. An additional calibrated photodiode allows to determine the incident photon flux.

To excite the N-K${\alpha}$  fluorescence radiation, an incident photon energy of $E_i = 680$  $\si{\eV}$ was chosen. Both angles (the incident angle $\theta_i$ and the azimuthal angle $\varphi_j$) were varied during the experiment. The measured fluorescence spectra were deconvoluted using the detector response functions for the fluorescence lines and the relevant background contributions.

The elemental mass depositions can be quantified in a reference-free approach\cite{Beckhoff2008} using the atomic fundamental parameters, describing the process of fluorescence production and the known instrumental parameters (e.g., the incident photon flux and the solid angle of detection)\cite{Hoenicke2019}.

\section{Simulation of fluorescence intensities}
\label{sec:sections}

The angular dependent GIXRF signals are directly related to the XSW field intensity distribution inside the nanostructure. The XSW field intensity distribution at each set of $\theta_i$ and $\varphi_j$ depends on both the dimensional and structural composition of the nanostructure. For a determination of these characteristics from the experimental data, we employ an FEM-based calculation of the XSW as already demonstrated by Soltwisch {\it et al.}~\cite{Soltwisch2018a}. Here, we use a rigorous Maxwell solver with a higher-order finite-element method provided by the JCMsuite package~\cite{pomplun_adaptive_2007}. Periodic boundary conditions are applied to both edges in the $x$ direction in agreement with the assumption of a periodic grating structure. The incoming electromagnetic wave is perpendicular to the x and y direction since in the GIXRF experiment the lines were oriented parallel to the incoming photon beam.


\begin{figure}[tbp]
\centering
\includegraphics[width=0.85\textwidth]{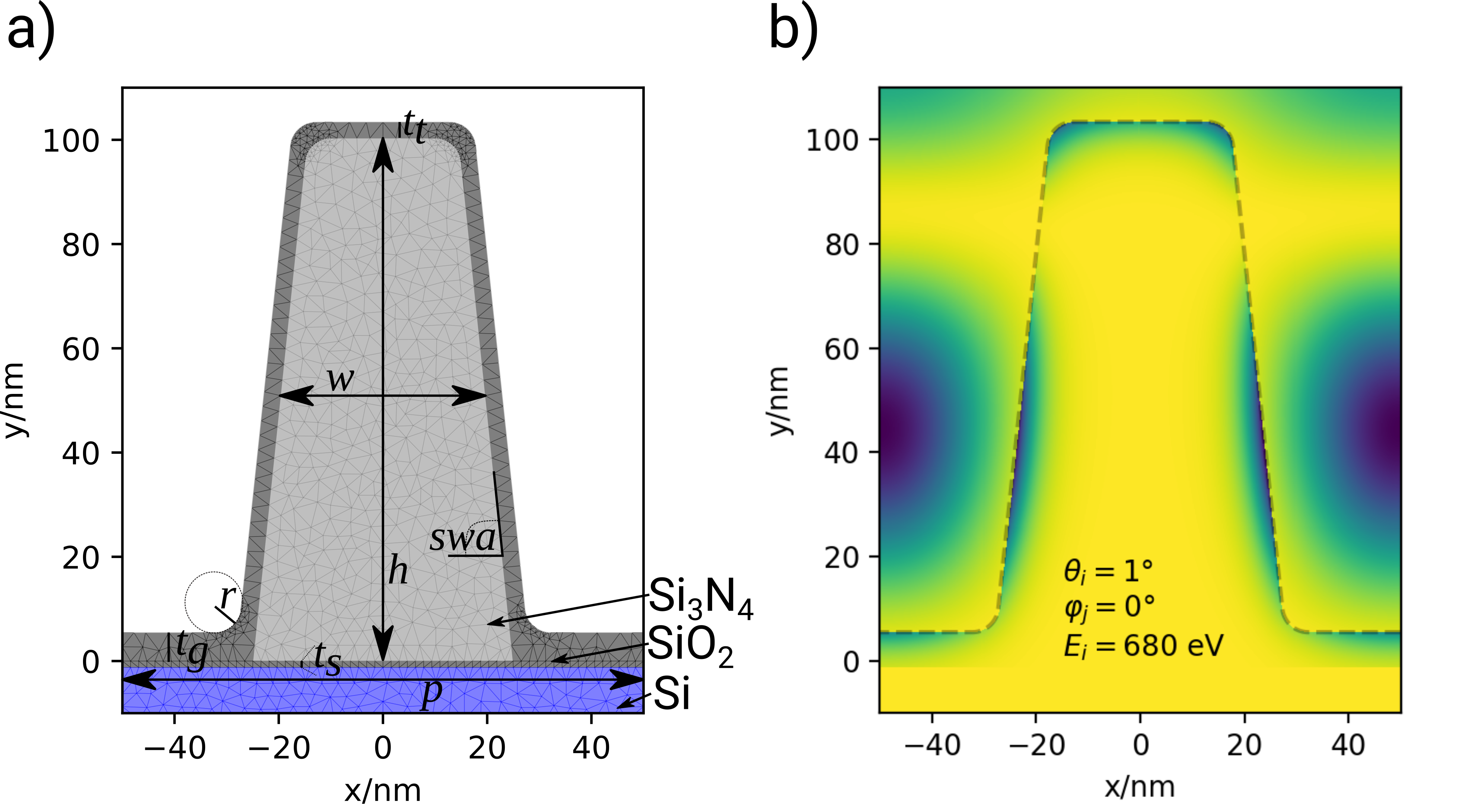}
	\caption{a) Cross-section with the finite-element mesh-grid showing the layout used for the simulation. The pitch $p$, the radius of the corner-rounding $r$, and the oxide layer on the substrate $t_s$ are kept constant during the optimization. The height $h$, the width $w$, the side-wall-angle $swa$, the oxide layer in the groove $t_g$ and the oxide layer of the grating $t_t$ were optimized as independent parameters. b) The calculated electric field strength in false color for a given orientation of the sample with respect to the incident beam. The field in the Si$_3$N$_4$ domain is multiplied by a factor to visually highlight the field distribution inside the sample.}
\label{fig:fem_mesh}
\end{figure}

For a conversion of the thereby calculated XSW field intensities within each material of the nanostructure to fluorescence intensities, we employ a modified Sherman equation\cite{Soltwisch2018a}.
We can use the calculated electric near-field intensity $|E(x,y)|^2$ distribution  inside the grating structure to extract a numerical approximation of the expected fluorescence intensity $F(\theta_i,E_i)$ per incident photon. For this purpose, we interpolate $|E(x,y)|^2$ inside a specific area to a Cartesian grid $(x,y)$ with sufficient discretization ($dx\times dy=1$ nm$^2$). To account for self-attenuation, every field intensity on this grid is damped with respect to the path length $y_{dis}=(y-y_0) / \cos(\theta_i)$ of the emitted fluorescence photons through the Si$_3$N$_4$ in the direction of the fluorescence detector. Both the solid angle $\Omega/4\pi$ and the detection efficiency for N-K$\alpha$ radiation $\epsilon_{E_f}$ are known for this detector. The attenuation coefficient $\mu_{E_i}$ for Si$_3$N$_4$ at the photon energy of the N-K$\alpha$ fluorescence line are taken from X-raylib\cite{T.Schoonjans2011} for bulk materials. $\rho$ is the density of the material determined from a X-ray reflectometry measurement at $E_i$. The mass fraction $W_i$ of nitrogen in Si$_3$N$_4$ is needed as well as $\tau(E_i)$ as the photo ionization cross section of the N-K shell \cite{T.Schoonjans2011} and $\omega_k$ as the fluorescence yield, which was determined experimentally like described for the determination of the O-K shell fluorescence by H\"onicke {\it et al.}\cite{P.Hoenicke2016a}

\begin{equation}
\frac{4\pi\sin\theta_i}{\Omega}\frac{F(\theta_i,\varphi_j,E_i)}{N_0\epsilon_{E_f}} =\frac{W_i \rho \tau(E_i) \omega_k}{\sum dx} \cdot \sum_{x}\sum_{y}  |E(x,y)|^2 \cdot \exp\left[-\rho\mu_{E_i}y_{dis}\right] dx dy\textrm{.} 
\label{eq:2Dsherman}
\end{equation}

Here, the normalized experimental data (left side of eq. \ref{eq:2Dsherman}) is reproduced by the 2D integration of the absolute electric field strength scaled by the atomic fundamental parameters describing the fluorescence production. In addition, self-attenuation of the fluorescence photons on their way to the detector is taken into account. Using this formula, the total emitted N-K$\alpha$ fluorescence intensity can be calculated from the electrical field distributions and compared to the normalized experimental data within an optimization algorithm such as particle swarm optimization (PSO) or Bayesian optimization (BO).
PSO is a stochastic global optimization method~\cite{Zhang2015,kennedy_particle_1995}. The particles of the swarm are moving randomly through the predefined parameter space. The velocity of each particle is guided by the particle's best
known position as well as the swarm's best
known position. The method is implemented based on the Python package pyswarm~\cite{pyswarm}.

Bayesian optimization (BO), in contrast, uses a stochastic model of an unknown objective function to be minimized in order to determine promising parameter values~\cite{shahriari2016taking,Garcia-Santiago2018}.
We use a Gaussian process (GP) as stochastic model, which is the most common choice. Given previous observations of the objective functions, a GP can predict the function value and its statistical uncertainty for each point of the parameter space $\mathbf{x}\in\mathcal{X}$ by means of GP regression -- a form of Bayesian inference. Based on this statistical information one can determine the expected improvement ${\rm EI}(\mathbf{x})=\mathbb{E}[{\rm max}(0,f_{\rm min} - f(\mathbf{x}))]$, i.e. the probabilistic expectation value of the one-sided difference ${\rm max}(0,f_{\rm min} - f(\mathbf{x}))$ between the function value $f(\mathbf{x})$ and the currently known lowest function value $f_{\rm min}$. The next sampling point is chosen at a position of maximized expected improvement. The numerical study is based on an in-house implementation of BO, which is integrated into the JCMsuite software package~\cite{JCMsuite}.

Since BO takes all previous function evaluations into account it can be more efficient than other heuristic global optimization strategies~\cite{Schneider2018} and local optimization strategies~\cite{schneider2019using}. For example, PSO computes the next sampling position solely based on the current position and velocity of one swarm member as well as based on the best seen position of the member and the swarm.

The error function \begin{equation}
\chi^2 = \sum_{\alpha_i,\varphi_j}{\frac{(I_{exp}(\alpha_i,\varphi_j) - I_{model}(\vec{gp},\alpha_i,\varphi_j))^2 }{\sigma_N^2(\alpha_i,\varphi_j)}}
\label{eq:chi}
\end{equation}
is minimized with respect to to 
5 different parameter $\vec{gp}$ (height $h$, width $w$, side-wall-angle $swa$, oxide layer in the groove $t_g$ and oxide layer on the grating $t_t$), see Fig.~\ref{fig:fem_mesh}.

\begin{figure}[tbp]
\centering
\includegraphics[width=1\textwidth]{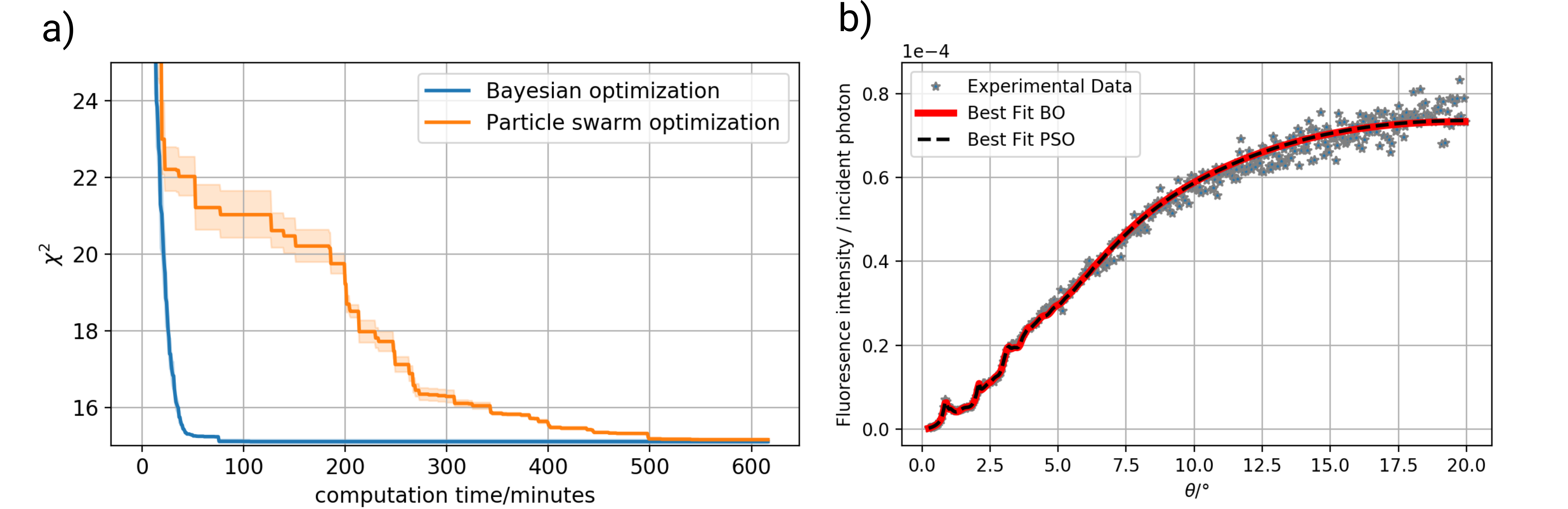}
	\caption{a) $\chi^2$ during the optimization process as a function of the total computation time of the optimization averaged over six independent optimization runs of Bayesian optimization and particle swarm optimization, respectively. The shading indicates the uncertainty of the average. Both optimization methods were configured to run six evaluations of the objective function in parallel. b) The best fit of the Bayesian optimization (red line) and the particle swarm optimization (dashed line) is compared to the measured N-K$\alpha$ fluorescence intensity.}
\label{fig:BO_vs_PSO}
\end{figure}

We have applied the two different global optimization approaches in order to benchmark their performance.
In Fig. ~\ref{fig:BO_vs_PSO} a) the improvement of the $\chi^2$ value as a function of the the total computation time for BO and PSO is shown. Due to the heuristic nature of the optimization methods, both optimization strategies were repeated 6 times and only the averaged values of $\chi^2$ are shown in the Figure. Clearly, BO requires a significantly shorter computation time than PSO to reach good results. On average, PSO takes five times longer to obtain similarly good fit values.

\section{Reconstruction of a grating from GIXRF measurements}

\begin{figure}[tbp]
\centering
\includegraphics[width=1\textwidth]{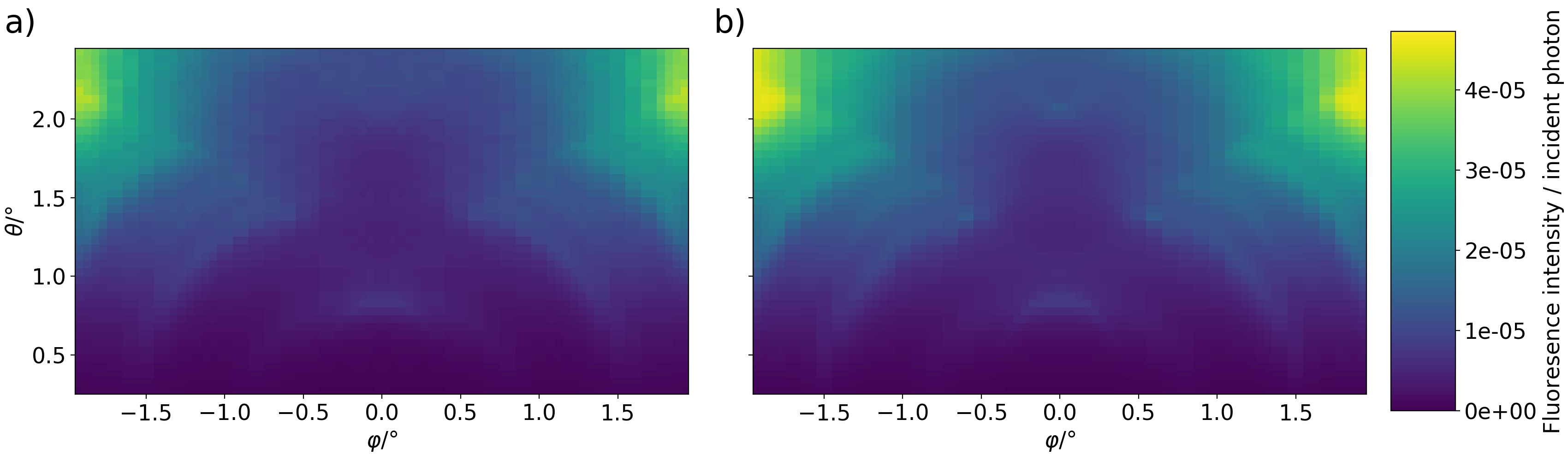}
	\caption{Comparison of the experimental data (a)) with the fit results of the Bayesian optician (b)) for different angle of incidence $\theta_i$ and $\varphi_j$}
\label{fig:maps}
\end{figure}

The best fit results obtained with both algorithms can be found in Fig. ~\ref{fig:BO_vs_PSO} b), which shows the measured fluorescence intensity per incident photon for increasing $\theta_i$ (grey stars) for $\varphi_j=0^{\circ}$. The variance of the data points for $\theta_i>5^{\circ}$ results from a reduced integration time of the spectra to reduce the overall measurement time. The modulation of the fluorescence curves for $\theta_i<5^{\circ}$ depends on the structure of the nanograting as already demonstrated by Soltwisch {\it et al.}~\cite{Soltwisch2018a} The red line shows the simulated XRF intensity for the results of the BO. The dashed black line shows the results for PSO. The described feature can also be found in both results and the difference between the results of the BO and the PSO are below 3\%. 
Fig. ~\ref{fig:maps} shows the comparison between the experimental data and the calculated intensity from the best fit of the BO for the full fluorescence intensity map for $\theta_i<2.5^{\circ}$ and $|\varphi_j|<2^{\circ}$. In both maps several similar features are visible. As it was shown in Fig. ~\ref{fig:fem_mesh} b), the main feature around $\theta_i=1^{\circ}$ and $\varphi_j=0^{\circ}$ originates from the strong confinement of the electric field inside the grooves. The overall agreement between the experimental data and the simulated data is good for the full angular regime. 

The obtained dimensional parameters of the grating are compared in table ~\ref{tab:results}. The slight differences between the results obtained by the BO and PSO can be explained with the slightly higher $\chi^2$ of the PSO compared to the $\chi^2$ of the BO. The particle swarm optimization was not able to find the global minimum in the given time. A tuning of the PSO parameters, like start velocities and weighting of the global best and local best solutions, could give better results. However, the results are in good agreement to the expected values of the grating. 

\begin{table}
\caption{The values of the geometrical parameters of the nominal nanostructure in comparison with the best fit of the Bayesian optimization and the best fit of the particle swarm optimization.}
\begin{tabular}{c|c|c|c}
geometrical parameter& nominal values & Bayesian optimization & particle-swarm optimization \\ 
\hline
height & $90$ nm  & $89.2$ nm & $90.8$ nm \\
width& $40$ nm& $43.9$ nm & $43.9$ nm \\
side-wall-angle& & $82.0^{\circ}$ & $82.4^{\circ}$ \\
oxide layer in the grating & & $2.7$ nm & $2.7$ nm \\
oxide layer of the groove & & $2.5$ nm & $2.6$ nm 
    
\label{tab:results}
\end{tabular}
\end{table}


\section{Conclusions}
In this work, we have shown that reference-free GIXRF is a promising experimental technique for the dimensional and elemental characterization of well ordered nanostructures. We demonstrate that GIXRF is able to simultaneously determine the composition and dimensional parameters of a 2D nanostructure with high sensitivity by quantitatively modeling the N-K$\alpha$ fluorescence intensity using a finite-element Maxwell solver. The reconstruction results obtained with Bayesian optimization or particle swarm optimization fit well to the experimental data and the results of the two different optimization methods are both close to each other as well as in good agreement to the nominal values. In this example, we demonstrate that Bayesian optimization converges much faster than classical heuristic optimization strategies. The method thus has a great potential for the reconstruction of even more complex 2D or 3D geometric shapes which require an enormous numerical effort.

\acknowledgments 
 
This project has received funding from the Electronic Component Systems for European Leadership Joint Undertaking under grant agreement No 783247 – TAPES3.
This Joint Undertaking receives support from the European Union's Horizon 2020 research and innovation programme and Netherlands, France, Belgium, Germany, Czech Republic, Austria, Hungary, Israel.
  
\bibliography{report} 
\bibliographystyle{spiebib} 

\end{document}